# A new method to derive the calculation formula of antenna total isotropic sensitivity

XiaoYu Hu
*China Academy of Information and Communications Technology*
Beijing, China
huxiaoyu@caict.ac.cn

*Abstract*—In the OTA test of the antenna spatial performance of communication equipment, the total radiated power TRP and the total isotropic sensitivity TIS are two important characterization parameters, representing the overall transmission and reception performance of the antenna in space. In order to gain a deeper understanding of the physical meaning of these two parameters and the calculation method of TIS, this paper proposes a new method for deriving the TIS calculation formula. This method uses the reciprocity principle of the transmitting antenna and the receiving antenna. By comparing the transmitting process, two corresponding receiving antenna parameters are newly defined in the receiving process analysis: receiving efficiency and receiving intensity. Combining the relationship between the antenna pattern and the receiving intensity, the relationship formula between TIS and equivalent isotropic sensitivity EIS is obtained after a simple deduction. This formula is consistent with the calculation formula given in the reference, but the derivation method in the reference is to use the ideal omnidirectional antenna as a reference, calculate the received power at the receiver input, and combine factors such as the actual antenna gain to derive the TIS calculation formula; the method proposed in this paper is to first derive the expression of the receiving directivity coefficient by definition, introduce the receiving strength parameter, calculate the full-space receiving power at the antenna, and derive the TIS calculation formula. In comparison, the derivation process in this paper calculates the received power from the input end of the receiving antenna, so there is no need to consider the efficiency of the antenna itself. The received power calculation starts directly from the receiving intensity. The entire derivation process is intuitive and easy to understand, avoiding complex concepts and formulas.

## I. Introduction

The antenna's transceiver performance plays a critical role in the communication quality of communication equipment, especially for today's highly integrated communication products. The air interface performance of a communication device is usually expressed by OTA(Over the Air). The antenna transmission performance is a necessary condition to ensure the quality of the communication link, and is usually described by the total radiated power (TRP) . For the receiving performance, there is also a corresponding description parameter, the total isotropic sensitivity (TIS), which reflects the spatial overall average sensitivity of the antenna. A good TIS means that the device has a higher receiving sensitivity and a stronger weak signal reception capability.

The test standards and methods for TRP and TIS of different communication standards are mainly developed by organizations such as CTIA and 3GPP. The test can be completed using facilities such as microwave anechoic chambers or reverberation chambers. In the anechoic chamber test method, the effective isotropic radiated power (EIRP) and equivalent isotropic sensitivity (EIS) at each azimuth angle are first measured, and then the average value is obtained by integrating on the full space sphere, and finally the TRP and TIS values are obtained. CTIA has given the derivation method of TRP and TIS in the relevant standards (Appendix E of [1]), and the derivation of the relationship between TIS and EIS refers to the calculation formula of [2]. From another perspective, this paper introduces a new receiving antenna parameter based on the reciprocity principle of the transmitting and receiving antennas, and gives a simple method for deriving the relationship between TIS and EIS. This method is completely derived from the perspective of the correspondence between the transmitting and receiving antennas, so for the sake of clarity, the derivation and calculation of the transmitting antenna TRP is first introduced below.

## II. Calculation formula for antenna total radiated power TRP

The schematic diagram of antenna transmission is shown in Fig. 1, The transmitter is connected to the antenna through a feeder line. The function of the antenna is to transmit the signal sent by the transmitter as efficiently as possible.

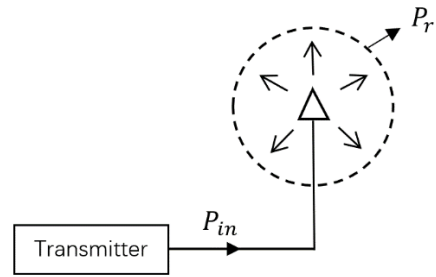

Fig. 1. Schematic diagram of antenna radiation process.

Assume that the input power of the antenna (i.e. the output power of the transmitter) is $P_{in}$ and the radiated power of the antenna is $P_r$, the relationship between the two is as follows:

$$P_r = P_{in} \cdot eff \qquad (1)$$

The $eff$ is the antenna radiation efficiency, and its value is between 0 and 1. For an ideal lossless antenna, the radiation efficiency is 1, that is, 100 % of the input power is radiated. Due to energy loss, the radiation efficiency of an actual antenna is always less than 100 %.

The relationship between antenna radiation efficiency $eff$ and antenna gain $G(\theta, \varphi)$ is,

$$G(\theta, \varphi) = D(\theta, \varphi) \cdot eff \qquad (2)$$

The $(\theta, \varphi)$ refers to the radiation angle of the radio wave in the spherical coordinate system with the center of the antenna as the origin, $G(\theta, \varphi)$ represents the antenna gain in the direction $(\theta, \varphi)$, $D(\theta, \varphi)$, The antenna directivity coefficient, represents the radiation ability of the antenna in the direction $(\theta, \varphi)$, and is defined as,

$$D(\theta, \varphi) = \frac{U(\theta, \varphi) \cdot 4\pi}{P_r} \quad (3)$$

Where $U(\theta, \varphi)$ is the radiation intensity of the antenna.

From (2), we can see that antenna gain consists of two parts: one part is the spatial directivity, which is usually called the antenna pattern; the other part is the energy conversion efficiency of the antenna, which is caused by the impedance mismatch and nonlinear factors of the circuit.

$P_r$ is the total radiated power TRP of the antenna, refers to the total power radiated by the antenna system to the surrounding space, which is defined as:

$$P_r = \oiint U(\theta, \varphi) \, d\Omega$$

$U(\theta, \varphi)$ is the antenna radiation intensity, which refers to the radiation power within a unit of solid angle. The total radiation power can be obtained by integrating $U(\theta, \varphi)$ within the range of $4\pi$ solid angles. Substituting $d\Omega = \sin\theta d\theta d\varphi$, we can get,

$$P_r = \iint U(\theta, \varphi) \sin\theta \, d\theta d\varphi \quad (4)$$

The equivalent isotropic radiated power $EiRP(\theta, \varphi)$ is defined as,

$$EiRP(\theta, \varphi) = P_{in} \cdot G(\theta, \varphi) \quad (5)$$

where $(\theta, \varphi)$ represents the azimuth in spherical coordinates and $G(\theta, \varphi)$ is the transmitting antenna gain,

$$G(\theta, \varphi) = \frac{U(\theta, \varphi) \cdot 4\pi}{P_{in}}$$

Substituting the above formula into (5) yields,

$$U(\theta, \varphi) = \frac{EiRP(\theta, \varphi)}{4\pi}$$

Substituting the above formula into TRP calculation formula (4) yields,

$$TRP = \frac{1}{4\pi} \iint EiRP(\theta, \varphi) \sin\theta \, d\theta d\varphi$$

Considering that the actual antenna's transmitted wave can be decomposed into two mutually perpendicular polarization directions, the above formula becomes,

$$TRP = \frac{1}{4\pi} \iint \left( EiRP_\theta(\theta, \varphi) + EiRP_\varphi(\theta, \varphi) \right) \sin\theta \, d\theta d\varphi$$

This is the formula for calculating $TRP$ in the Appendix E of [1], where $EiRP_\theta(\theta, \varphi)$ and $EiRP_\varphi(\theta, \varphi)$ represent the polarization components of the equivalent isotropic radiated power $EiRP(\theta, \varphi)$ in the θ direction and the φ direction respectively. In the actual antenna OTA anechoic chamber test, the equivalent isotropic radiated power $EiRP_\theta(\theta, \varphi)$ and $EiRP_\varphi(\theta, \varphi)$ of each point in space are first obtained, and then the total radiated power TRP is calculated using the above formula.

III. DERIVATION OF THE CALCULATION FORMULA FOR TOTAL ISOTROPIC SENSITIVITY TIS

The schematic diagram of the antenna receiving process is shown in Fig. 2, The antenna is connected to the receiver through a feeder, The function of the receiving antenna is to send all the wireless signals received in the space to the receiver for further processing.

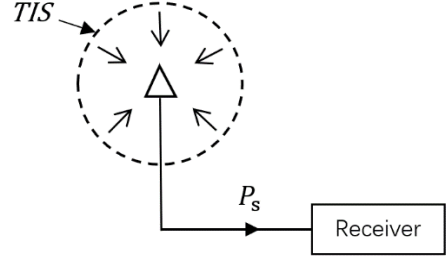

Fig. 2. Schematic diagram of antenna receiving process.

When the signal input power of the receiver is exactly equal to the receiver sensitivity $P_S$, according to the definition of TIS in [1], the total power received by the receiving antenna in space is TIS. Corresponding to the radiation efficiency of the transmitting antenna $eff$, a new receiving antenna parameter is introduced: receiving efficiency $eff'$, and the following relationship also holds,

$$P_S = TIS \cdot eff' \quad (6)$$

The value of $eff'$ is between 0 and 1, and for a lossless ideal antenna, the value is 1.

For the equivalent isotropic sensitivity EIS, it can be seen from its definition that,

$$EIS(\theta, \varphi) = \frac{P_S}{G_r(\theta, \varphi)} \quad (7)$$

Where $G_r(\theta, \varphi)$ is the receiving gain of the receiving antenna in the $(\theta, \varphi)$ direction.

Combining the above two equations, we can get,

$$TIS \cdot eff' = EIS(\theta, \varphi) \cdot G_r(\theta, \varphi) \quad (8)$$

Similar to the transmitting antenna, we also have

$$G_r(\theta, \varphi) = D_r(\theta, \varphi) \cdot eff'$$

substituting it into (8) and simplifying it to get,

$$D_r(\theta, \varphi) = \frac{TIS}{EIS(\theta, \varphi)} \quad (9)$$

Here, $D_r(\theta, \varphi)$ represents the receiving pattern of the antenna, which is essentially similar to its transmitting pattern $D(\theta, \varphi)$. When we consider the situation where the received signal strength is evenly distributed at each point in space and the total spatial received power value is TIS, referring to the definition of the transmitting directivity coefficient, the following relationship holds,

$$D_r(\theta,\varphi) = \frac{4\pi \cdot U'(\theta,\varphi)}{TIS} \tag{10}$$

The new parameter $U'(\theta,\varphi)$ introduced in the (10) corresponds to the antenna radiation intensity $U(\theta,\varphi)$, $U'(\theta,\varphi)$ can be called the antenna receiving intensity, It is the antenna receiving intensity when the power received by the receiver is just equal to the receiver sensitivity $P_s$.

Combining (9) and (10), we can get,

$$U'(\theta,\varphi) = \frac{TIS^2}{4\pi \cdot EIS(\theta,\varphi)}$$

The above formula also holds true for the components of $U'(\theta,\varphi)$ in the two polarization directions θ and φ of the electric field,

$$U'_\theta(\theta,\varphi) = \frac{TIS^2}{4\pi \cdot EIS_\theta(\theta,\varphi)}$$

$$U'_\varphi(\theta,\varphi) = \frac{TIS^2}{4\pi \cdot EIS_\varphi(\theta,\varphi)}$$

$U'(\theta,\varphi)$ is a power scalar, which is composed of the sum of the polarization components of the electric field intensity in the θ and φ directions, and $U'(\theta,\varphi)$ can be expressed as

$$U'(\theta,\varphi) = U'_\theta(\theta,\varphi) + U'_\varphi(\theta,\varphi)$$

$$= \frac{TIS^2}{4\pi \cdot EIS_\theta(\theta,\varphi)} + \frac{TIS^2}{4\pi \cdot EIS_\varphi(\theta,\varphi)} \tag{11}$$

Because $U'(\theta,\varphi)$ is defined as the receiving intensity, integrating it within the $4\pi$ solid angle range gives the total receiving power. For the sensitivity we consider, that is TIS,

$$TIS = \iint U'(\theta,\varphi) \sin\theta \, d\theta d\varphi$$

Substituting $U'(\theta,\varphi)$ expression (11) into the above formula, we get,

$$TIS = \iint \left[ \frac{TIS^2}{4\pi \cdot EIS_\theta(\theta,\varphi)} + \frac{TIS^2}{4\pi \cdot EIS_\varphi(\theta,\varphi)} \right] \sin\theta \, d\theta d\varphi$$

Subtracting TIS from both sides,

$$\frac{1}{TIS} = \frac{1}{4\pi} \iint \left[ \frac{1}{EIS_\theta(\theta,\varphi)} + \frac{1}{EIS_\varphi(\theta,\varphi)} \right] \sin\theta \, d\theta d\varphi$$

Or written as:

$$TIS = \frac{4\pi}{\iint \left[ \frac{1}{EIS_\theta(\theta,\varphi)} + \frac{1}{EIS_\varphi(\theta,\varphi)} \right] \sin\theta \, d\theta d\varphi}$$

The above formula is the calculation formula of TIS, which is consistent with the derivation result of TIS in reference [1]. In actual testing, the EIS of the device under test at each point in space is first measured, and then the TIS can be calculated using the above formula.

In the Appendix E.1 of reference [1], the derivation method of the TIS formula is to use an ideal omnidirectional antenna as a reference, calculate the received power at the receiver input, then derive the TIS calculation formula. The method proposed in this paper first obtains $D_r(\theta,\varphi)$ from the definition, introduces the receiving strength parameter, calculates the full-space receiving power at the antenna, and derives the TIS calculation formula. In comparison, the derivation process of this paper calculates the received power from the input of the receiving antenna, so there is no need to consider the efficiency of the antenna itself. The received power calculation starts directly from the receiving strength. The physical concept is relatively clear and helps to understand the antenna efficiency and parameters such as EIS and TIS from the perspective of reciprocity of transmission and reception.

From the above derivation process, it can be seen that the total received power of the antenna in space becomes the received power of the receiver after a certain amount of self-energy loss; this process is just like the output power of the transmitter, which becomes the antenna radiation power after a certain amount of internal loss. When the received power of the receiver is exactly the receiver sensitivity $P_s$, the corresponding antenna space received power is TIS. Therefore, except for the ideal lossless antenna, the TIS of the actual antenna is always greater $P_s$ than. $P_s$ is the conducted sensitivity, which has nothing to do with the antenna performance; TIS is the spatial or radiated receiving sensitivity, which reflects the total receiving sensitivity of the device including the antenna. EIS is the received power of the equivalent ideal lossless omnidirectional antenna in a certain direction. It can be seen from (7) that when $P_s$ is constant, EIS is only related to the receiving gain of the antenna. When the receiving gain of the antenna in a certain direction is greater than 1, the antenna plays a signal amplification role in this direction, and the receiver sensitivity $P_s$ is greater than EIS; on the contrary, if the receiving gain is less than 1, the received signal is attenuated and the receiver sensitivity $P_s$ is less than EIS.

IV. CONCLUSION

OTA test of wireless communication equipment, whether it is mobile communication or Wi-Fi, Bluetooth and other technologies, TRP and TIS are two very important basic parameters. Compared with the total transmit power TRP, the concept and calculation formula of total isotropic sensitivity TIS are relatively complicated. Starting from the principle of antenna reciprocity, this paper defines two new parameters of receiving antenna: receiving efficiency $eff'$ and receiving intensity $U'(\theta,\varphi)$ by comparing the radiation efficiency and radiation intensity parameters of the transmitting antenna system. Combining the relationship between the antenna pattern and the receiving intensity, the calculation formula of the relationship between TIS and EIS is derived. It is hoped that the parameters and methods introduced in this paper can help to deepen the understanding of the antenna receiving parameter TIS and can be applied in the actual system analysis and calculation.